\documentclass[12pt]{article}
\usepackage{amsmath}
\usepackage[cp1251]{inputenc}
\usepackage{bm}
\usepackage{color}
\usepackage{caption}
\usepackage{subcaption}
\usepackage{graphicx}
\usepackage{braket}
\begin{document}
\title{Relation of curvature and torsion of weighted graph states  with graph properties and its studies on a quantum computer}

\author{
Kh. P. Gnatenko$^1$\footnote{khrystyna.gnatenko@gmail.com} \\
$^1$Professor Ivan Vakarchuk Department for Theoretical Physics,\\
Ivan Franko National University of Lviv,\\
12, Drahomanov St., Lviv, 79005, Ukraine.\\
}

\maketitle

\begin{abstract}
Quantum states of spin systems that can be represented with weighted graphs $G(V, E)$ are studied. The velocity, curvature, and torsion of these states are examined. We find that the velocity of quantum evolution is determined by the sum of the weighted degrees of the nodes in the graph, constructed by raising to the second power the weights of graph $G(V, E)$. The curvature depends on the sum of the weighted degrees of nodes in graphs constructed by raising the weights to the second and fourth powers. It also depends on the sum of the products of the weights of edges forming squares in graph $G(V, E)$. The torsion is related to the sum of the weighted degrees of nodes in graphs constructed by raising the weights to the second, third, and fourth powers, as well as the sum of the products of the weights of edges in graph $G(V, E)$ forming triangles $S_3$. Geometric properties of quantum graph states and the sum of the weighted degrees of nodes have been calculated with quantum programming on IBM's quantum computer for the case of a spin chain.

\end{abstract}

\vspace{0.5cm}

\section{Introduction}

Quantum analogs of well-known geometric characteristics of classical trajectories, namely the curvature and torsion of evolutionary quantum states, were introduced in \cite{Brody96, Lab17}. The curvature of quantum evolution indicates the deviation of the evolutionary quantum state vector from the geodesic line (for details, see \cite{Lab17}). Torsion relates to the deviation of the evolutionary quantum state vector from the plane of evolution (see \cite{Lab17}). In \cite{Brody96}, the expression for the curvature of quantum evolution was derived based on studies of the geometry of quantum statistical interference. In \cite{Lab17}, the relationship between the curvature and torsion of evolutionary quantum states and energy fluctuations was established. It is worth mentioning that geometric ideas play a significant role in the study of quantum systems and their evolution, quantifying the entanglement of quantum states (the geometric measure of entanglement), see, for instance, \cite{Shim95, Brody01, Brody07, Fry17, Anandan90, Abe93, Grigorenko91, Mosta07, Bender08, Kuzmak16,  Fryd08, Kuzmak15} and references therein.

In the present paper, we study the geometric properties of weighted graph states. These are multi-qubit quantum states that can be represented using graphs with edges characterized by weights (weighted graphs). It is worth noting that quantum graph states are entangled states that have been intensively studied \cite{Markham, Schlingemann, Bell, Wang, Mooney, Mazurek, Shettell, Hein, Guhne, Qian, Vesperini1, Vesperini2}. The properties of quantum graph states are related to the properties of the corresponding graphs. In the case of unweighted graph states, this relationship was demonstrated, for instance, in \cite{Laba22, Gna21}. These states appear in various quantum information problems, such as quantum cryptography \cite{Markham, Qian}, quantum error correction algorithms \cite{Schlingemann, Bell, Mazurek}, quantum machine learning \cite{Zoufal, Gao}, and others

Quantum graph states can be constructed with action of two-qubit gates on an initial separable multi-qubit quantum state. In this case, qubits are represented as vertices in a graph, and the action of a two-qubit gate corresponds to linking two vertices with an edge. Many papers are devoted to the study of quantum graph states constructed with action of controlled-Z gates \cite{Wang, Mooney, Alba, Mezher, Akhound, Haddadi, Cabello}. Also, the graph states constructed with controlled phase gates $CP_{ij}$ \cite{Susulovska, Susulovska1},
$RXX$ gates \cite{Gna2024} were examined and the entanglement of the states was quantified with quantum calculations.

Quantum graph states of spin systems with the Ising model corresponding to unweighted graphs were studied in \cite{Gna21, Laba22}. In \cite{Gna21}, the geometric measure of entanglement of the states was examined. It was found that the entanglement of a spin in the graph state is related to the degree of the corresponding vertex in the graph.

In \cite{Laba22}, the geometric characteristics of evolutionary quantum graph states corresponding to unweighted graphs were considered, and their relation with the number of triangles, squares, and edges in the graph was found \cite{Laba22}.  Also there similar calculation where done in \cite{Vesperini1}.

In the present paper, we study quantum states corresponding to spin systems described by the Ising model with anisotropic interactions. These states are weighted graph states and can be represented with weighted graphs. We examine the velocity, curvature, and torsion of these states. For this purpose, we use the relation of these properties with the fluctuations of energy obtained in \cite{Lab17}. We find a relationship between the geometric properties of the graph states and the sum of the products of weights of edges forming triangles and squares in the graph, as well as the sum of the weighted degrees of nodes in the graphs constructed by raising the weights of the graph
$G(V,E)$ to the second, third, and fourth powers. The curvature and torsion are calculated with quantum programming on IBM's quantum computer    $\textrm{ibm\_sherbrooke}$ \cite{kk} for particular case of quantum graph state corresponding to a chain.

The structure of the paper is as follows: In Section 2, the velocity, curvature, and torsion of weighted graph states are calculated analytically based on their dependence on the fluctuations of energy. The relationship between the geometrical characteristics and the graph properties is obtained. Section 3 is devoted to the studies of the velocity, curvature, and torsion of weighted graph states using quantum computing. Results of quantum calculations for curvature and torsion  on IBM's quantum computer are presented.Conclusions are presented in Section 4.

\section{Geometric characteristics of evolutionary weighted graph states }

Let us consider a system described by the Ising model with  Hamiltonian
\begin{eqnarray}
H_I={1\over 2}\sum_{i,j}J_{ij}\sigma^z_i\sigma^z_j,\label{ising}
\end{eqnarray}
where $\sigma_i^z$ is the Pauli matrix that corresponds to spin $i$. Constants $J_{ij}$ are interaction couplings.
The evolutionary state of the system
\begin{eqnarray}
\ket{\psi(t)}=\exp\left(-\frac{iH_It}{\hbar}\right)\ket{\psi_0} \label{evol1}
\end{eqnarray}
can be considered as a quantum graph state
\begin{eqnarray}
\ket{\psi_G}=\prod_{(i,j) \in E} RZZ_{ij}(\phi_{ij}) \ket{\psi_0}. \label{graph_state}
\end{eqnarray}
In (\ref{evol1}), (\ref{graph_state}),
$\ket{\psi_0}$ is an initial state,  the $RZZ$ gate is defined as  $RZZ(\phi_{ij})=\exp(-i\phi_{ij}\sigma^z_i\sigma^z_j/2)$, and $\phi_{ij}=2J_{ij}t/\hbar$.
State (\ref{graph_state}) can be represented with an weighted graph $G(V, E)$ with vertices $V$ corresponding to spins and edges $E$ illustrating interactions between them. Constant $J_{ij}$ corresponds to the weight of the edge $(i,j)$ and can be considered as an element of the adjacency matrix $A$ of a graph $G(V, E)$.

Let us find the geometric characteristics of the weighted graph states and study their relations with the graph properties.
According to the results of paper \cite{Lab17} these characteristics are related with the fluctuations of energy $\langle\Delta H_I^2\rangle$
$\langle\Delta H_I^3\rangle$,
$\langle\Delta H_I^4\rangle$.  Hamiltonian under consideration  (\ref{ising}) does not depend on time. So, to study $\langle\Delta H_I^n\rangle$ ($n=2,3,4$) we can  calculate $\bra{\psi_0}\Delta H_I^n\ket{\psi_0}$.

We consider the initial state
\begin{eqnarray}
\ket{\psi_0}=\ket{++..+}, \label{psi0}
\end{eqnarray}
where $\ket+=(\ket0+\ket1)/\sqrt{2}$.
Taking into account $\bra{\psi_0} H_I \ket{\psi_0}=0$, we have
$\langle\Delta H_I^n\rangle=\bra{\psi_0} \Delta H_I^n \ket{\psi_0}=\bra{\psi_0}  H_I^n \ket{\psi_0}$.

Let us calculate $\langle \Delta H_I^2\rangle$. We can write
\begin{eqnarray}
\langle \Delta H_I^2\rangle
={1\over 4}\sum_{i_1,j_1}\sum_{i_2,j_2}J_{i_1j_1}J_{i_2j_2}\bra{\psi_0}
\sigma^z_{i_1}\sigma^z_{j_1}\sigma^z_{i_2}\sigma^z_{j_2}\ket{\psi_0}=\nonumber\\={1\over 2}\sum_{i,j}J^2_{ij}={1\over 2}\sum_{i}(A^2)_{ii}.
\end{eqnarray}
Here $(A^2)_{ii}$ are diagonal elements of squared adjacency matrix $A$, $\ket{\psi_0}$ is given by (\ref{psi0}).
Let us consider a graph $G^{(2)}(V,E)$ constructed by squaring the weights of corresponding edges in $G(V,E)$. Note that
$(A^2)_{ii}=\sum_jJ^2_{ij}=n^{(2)}_i$, $n_i^{(2)}$ is the weighted degree of node $i$ (sum of the weights of edges incident to node $i$) in graph $G^{(2)}(V,E)$.  Therefore,  we can also write
\begin{eqnarray}
\langle \Delta H_I^2\rangle=\frac{1}{2}\sum_i n^{(2)}_i.\label{h21}
\end{eqnarray}
So, quadratic fluctuations of energy are related to the sum of weighted degrees of nodes in graph $G^{(2)}(V,E)$ characterized by the adjacency matrix with elements $J_{ij}^2$.

The velocity of quantum evolution (\ref{velocity}) is related to the quadratic fluctuations of energy. It reads
\begin{eqnarray} \label{velocity}
v={ds\over dt}=\frac{\gamma\sqrt{\braket{(\Delta H_I)^2}}}{ \hbar},
\end{eqnarray}
where $\gamma$ is a constant, see \cite{Anandan90}. For spin system with Ising model, taking into account  (\ref{h21}), we can write
\begin{eqnarray} \label{velocity1}
v=\frac{\gamma}{ \hbar}\sqrt{\frac{\sum_i n^{(2)}_i}{2}}.
\end{eqnarray}
So, the velocity of evolution is also related to the sum of weighted degrees of nodes in graph $G^{(2)}(V,E)$.

Let us examine $\langle \Delta H_I^3\rangle$. For Hamiltonian (\ref{ising}) We can write
\begin{eqnarray}
\langle \Delta H_I^3\rangle=\frac{1}{8}\sum_{i_1,j_1}\sum_{i_2,j_2}\sum_{i_3,j_3}
J_{i_1j_1}J_{i_2j_2}J_{i_3j_3}\times\nonumber\\ \times \bra{\psi_0}
\sigma^z_{i_1}\sigma^z_{j_1}\sigma^z_{i_2}\sigma^z_{j_2}\sigma^z_{i_3}\sigma^z_{j_3}\ket{\psi_0}=\frac{3!}{6}\sum_{i,j,k}J_{ij}J_{jk}J_{ki}=\frac{3!}{6}\sum_{i}(A^3)_{ii},\label{h31}
\end{eqnarray}
where $(A^3)_{ii}$ are diagonal elements of cubed adjacency matrix $A^3$, number $3!$ is
the number of combinations of three edges,  $\ket{\psi_0}$ is given by (\ref{psi0}).  Note that
\begin{eqnarray}
S_3=\frac{1}{6}\sum_{i,j,k}J_{ij}J_{jk}J_{ki}
\end{eqnarray}
is the sum of products of weights of these  edges $(i,j)$, $(j,k)$, $(k,i)$ in $G(V,E)$  creating triangle (multiplier $1/6$  is present because in the sum each triangle is accounted $6$ times). So, we have
\begin{eqnarray}
\langle \Delta H_I^3\rangle=3!S_3.\label{h32}
\end{eqnarray}

So, for cubic fluctuations of energy  we have obtained that they are related to  the sum of products of weights of edges creating triangles in graph $G(V,E)$ (\ref{h32}).
In the case of $J_{ij}=J$, $S_3=J^3n_3$, where $n_3$ is the number of triangles in $G(V,E)$.
So, the expression for $\langle \Delta H_I^3\rangle$ is reduced to $\langle \Delta H_I^3\rangle=3!J^3n_3$, that was obtained in \cite{Laba22}.

Let us also calculate $\langle\Delta H_I^4\rangle$. We have
\begin{eqnarray}
\langle\Delta H_I^4\rangle=
\frac{1}{16}\sum_{i_1, j_1}\sum_{i_2, j_2}\sum_{i_3, j_3}\sum_{i_4, j_4} J_{i_1j_1}J_{i_2j_2}J_{i_3j_3}J_{i_4j_4}\times \nonumber\\ \times \bra{\psi_0}\sigma^z_{i_1}\sigma^z_{j_1}\sigma^z_{i_2}\sigma^z_{j_2}\sigma^z_{i_3}\sigma^z_{j_3}\sigma^z_{i_4}\sigma^z_{j_4}\ket{\psi_0}=\nonumber\\=
\frac{4!}{8}\sum_{i,j,k,l}J_{ij}J_{jk}J_{kl}J_{li}(1-\delta_{ik})(1-\delta_{jl})+\nonumber\\+\frac{3}{4}\sum_{i,j,k,l}J^2_{ij}J^2_{kl}(1-\delta_{ik}\delta_{jl})(1-\delta_{il}\delta_{jk})+\frac{1}{2}\sum_{i,j}J^4_{ij},\label{dh4}
\end{eqnarray}
  We use notation
\begin{eqnarray}
S_4=\frac{1}{8}{\sum_{i,j,k,l}}_{\substack{ \hspace{0.10cm}i\neq k,\\ j\neq l}}J_{ij}J_{jk}J_{kl}J_{li}
\end{eqnarray}
for  the sum of  products of weights of four edges  $(i,j)$, $(j,k)$, $(k,l)$, $(l,i)$ in graph $G(V,E)$  that create square, $J_{ij}J_{jk}J_{kl}J_{li}$, $i\neq k$, and $j\neq l$. In the sum each square is accounted $8$ times, so multiplier $1/8$  is present.   To obtain  (\ref{dh4}), we take into account that the expression
$\bra{\psi_0}\sigma^z_{i_1}\sigma^z_{j_1}\sigma^z_{i_2}\sigma^z_{j_2}\sigma^z_{i_3}\sigma^z_{j_3}\sigma^z_{i_4}\sigma^z_{j_4}\ket{\psi_0}$
is equal to
1 if the edges characterized by $J_{i_1j_1}$, $J_{i_2j_2}$, $J_{i_3j_3}$, $J_{i_4j_4}$
 form a square in the graph (see the first term in (\ref{dh4}), the multiplier
$4!$ represents the number of permutations of the four edges); if two edges that connect different vertices in the graph are multiplied twice (see the second term in (\ref{dh4})), or if one edge with $J_{ij}$  is multiplied four times (see the third term in (\ref{dh4})), the corresponding multipliers between the terms are found by calculating all combinations of these cases.

The obtained result (\ref{dh4}) can be rewritten as follows
\begin{eqnarray}
\langle\Delta H_I^4\rangle=
4!S_4+\frac{3}{4}(\sum_i n^{(2)}_i)^2-\sum_i n^{(4)}_i,\label{dh41}
\end{eqnarray}
where $n^{(4)}_i=\sum_{j}J^4_{ij}$ is the weighted degree of $i$-th node in graph $G^{(4)}(V,E)$, constructed raising to the fourth the weights of corresponding edges in $G(V,E)$ (the elements of adjacency matrix of $G^{(4)}(V,E)$ are $J^4_{ij}$).

Note, that in the case of $J_{ij}=J$, we have $S_4=k_4J^4$
where $k_4$ is the number of squares in graph $G(V,E)$. Also, if graph is unweighted ($J_{ij}=J$)
$\sum_{i}n^{(2)}_i=\sum_{i,j}J^2_{ij}=2J^2k_2$,  $\sum_{i}n^{(4)}_i=\sum_{i,j}J^4_{ij}=2J^4k_2$, $k_2$ is the number of edges in $G(V,E)$. So,  the result (\ref{dh41}) reduces to that obtained in \cite{Laba22}. Mean value  $\langle\Delta H^4\rangle$ depends on the number of squares $k_4$ and number of edges $k_2$ in unweighted graph $G(V,E)$, $\langle\Delta H_I^4\rangle=
J^4\left(k_2+3k_2(k_2-1)+4!k_4\right)$
\cite{Laba22}.

In \cite{Lab17} it was found that the curvature is related to fluctuation of the energy as
\begin{eqnarray}
{\gamma^2\over R^2}={\langle(\Delta
H_I)^4\rangle-\langle(\Delta H_I)^2\rangle^2\over\langle(\Delta
H_I)^2\rangle^2},\label{curvvv}
\end{eqnarray}
Using the obtained results for $\langle\Delta H_I^2\rangle$,  $\langle\Delta H_I^4\rangle$ and taking into account  expression for curvature (\ref{curvvv})  we find
\begin{eqnarray}
{\gamma^2\over R^2}={\bar\kappa}=
\frac{96 S_4+2(\sum_i n^{(2)}_i)^2-4\sum_i n^{(4)}_i}{(\sum_i n^{(2)}_i)^2}.\label{res2}
\end{eqnarray}
 So, curvature (\ref{res2}) is related to the sum of the weighted degrees of nodes in the graphs $G^{(2)}(V,E)$, $G^{(4)}(V,E)$ and the sum of products  of weights of edges creating squares in graph $G(V,E)$.

 The torsion of quantum evolution depends on the fluctuations of energy as
\begin{eqnarray}\label{bartor}
 \bar\tau=
{\langle(\Delta H_I)^4\rangle-\langle(\Delta
H_I)^2\rangle^2\over\langle(\Delta H_I)^2\rangle^2}-{\langle(\Delta
H_I)^3\rangle^2\over\langle(\Delta H_I)^2\rangle^3},
\end{eqnarray}
see \cite{Lab17}.
 On the basis of (\ref{h32}), (\ref{dh4}), (\ref{bartor}) for torsion  we obtain
\begin{eqnarray}
\bar\tau=
\frac{96 S_4+2(\sum_i n^{(2)}_i)^2-4\sum_i n^{(4)}_i}{(\sum_i n^{(2)}_i)^2}-\frac{288S_3^2}{(\sum_i n^{(2)}_i)^3}.
\label{res3}
\end{eqnarray}

So, torsion is determined by  the sums of the weighted degrees of nodes in the graphs $G^{(2)}(V,E)$, $G^{(3)}(V,E)$, $G^{(4)}(V,E)$. It is also related to the sum of products  of weights of edges creating triangles $S_3$ and the sum of products  of weights of edges creating squares $S_4$ in graph $G(V,E)$.

In the next section the obtained results will be used for the construction of the quantum protocols for studies of the geometrical properties of weighted graph states with quantum computing.

\section{Study of the geometrical properties of weighted graph states with quantum programming}

According to the result (\ref{h21}) quadratic fluctuations of energy in graph states corresponding to the weighted graphs are related with the sum of weighted degrees of nodes of graph $G^{(2)}(V,E)$. Also, the velocity of evolution (\ref{velocity1}) is determined by the value of $\langle\Delta H_I^2\rangle$. So, studying $\langle\Delta H_I^2\rangle$ on a quantum device we can determine the sum of weighted degrees of nodes of graph $G^{(2)}(V,E)$ and find velocity of evolution.

The value of $\langle\Delta H_I^2\rangle$ can be quantified with quantum programming on the basis of studies of the mean value of the evolution operator at small times. For small times we can write
\begin{eqnarray}
|\langle U \rangle|^2=|\bra{\psi_0}\exp\left(\frac{-iH_It}{\hbar}\right)\ket{\psi_0}|^2=1-\frac{\braket{\Delta H_I^2}}{\hbar^2}t^2.\label{rel}
\end{eqnarray}
So, studying time dependence of $|\langle U \rangle|^2$ with quantum programming one finds $\braket{\Delta H_I^2}$ and therefore the sum of weighted degrees of nodes of graph $G^{(2)}(V,E)$ (\ref{h21}), and the velocity of evolution (\ref{velocity1}).

\begin{figure}[h!]
		\centering
\includegraphics[scale=0.6]{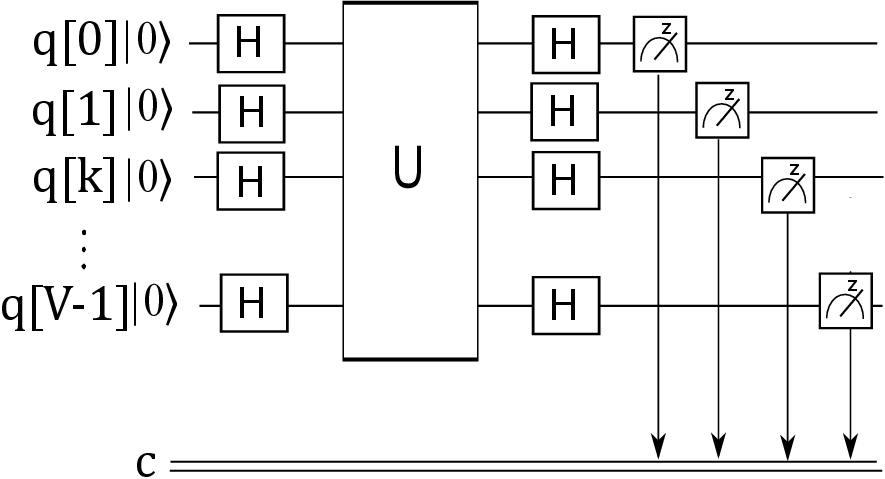}
		\caption{ Quantum protocol for quantifying $|\langle U \rangle|^2$ with quantum programming.}
		\label{fig1}
\end{figure}

Quantum protocol for studies of the mean value of evolution operator  is presented in Fig. \ref{fig1}.  In the protocol $U$ is the operator of evolution. In  the case of spin system with Ising model (\ref{ising}) this operator can be represented with $RZZ$ gates.
The value $|\langle U \rangle|^2$ can be determined on the basis of the results of measurements in the standard basis. We have $|\langle U \rangle|^2=|\langle{00..0}|{\tilde{\psi}}\rangle|^2$, where $\ket{\tilde{\psi}}=H^{[V]}UH^{[V]}\ket{00..0}$, $H^{[V]}=\prod^{V-1}_{i=0}H_i$, $H_i$ is Hadamard gate acting on qubit $q[i]$.

\begin{figure}[h!]
		\centering
\includegraphics[scale=0.6]{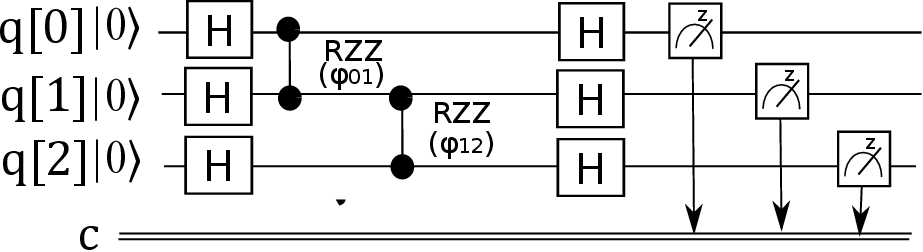}
		\caption{ Quantum protocol for measuring $|\langle U \rangle|^2$ with quantum programming for spin system with Hamiltonian (\ref{hch}).}
		\label{fig11}

  \end{figure}

Let us consider particular case of spin system. We examine a spin chain with the following Hamiltonian

\begin{eqnarray}
H_{ch}=J_{01}\sigma^z_0\sigma^z_1+J_{12}\sigma^z_1\sigma^z_2,\label{hch}
\end{eqnarray}
Weighted graph state
\begin{eqnarray}
\ket{\psi(t)}=\exp\left(-\frac{itJ_{01}\sigma^z_0\sigma^z_1}{\hbar}\right)\exp\left(-\frac{itJ_{12}\sigma^z_1\sigma^z_2}{\hbar}\right)\ket{+++} \label{evol}
\end{eqnarray}
can be represented with a chain with three vertices $0,1,2$ and two edges $(0,1)$ $(1,2)$ characterized by weights $J_{01}$, $J_{12}$.
Let us find $\langle\Delta H_{ch}^2\rangle$ with quantum programming. For this purpose we construct quantum protocol presented in Fig. \ref{fig11}. In the protocol parameters $\phi_{ij}$ are related with $J_{ij}$ and time. They read $\phi_{ij}=2J_{ij}t/\hbar$.
We consider $J_{01}=J$, $J_{12}=2J$. In this case we have  $\phi_{01}=2Jt/\hbar=2\phi$ and  $\phi_{12}=4Jt/\hbar=4\phi$, where for convenience we consider notation $\phi=Jt/\hbar$.

We run the protocol Fig. \ref{fig11} on IBM's quantum computer    $\textrm{ibm\_sherbrooke}$ (qubits $Q_1$, $Q_2$, $Q_3$)  for small times. Namely we consider parameter $\phi$ in range from  $-3\pi/32$ to $3\pi/32$ changing with step $\pi/64$, and number of shots equals $1024$. The calibration parameters of IBM's quantum computer  $\textrm{ibm\_sherbrooke}$ are shown in Table \ref{trr1}.

 \begin{table}[!!h]
\caption{The calibration parameters of IBM's quantum computer  $\textrm{ibm\_sherbrooke}$   on 1 August 2024 \cite{kk}.}\label{trr1}
\begin{tabular}{ c c c c }
       & $Q_1$ &  $Q_2$ & $Q_3$ \\
Readout error   ($10^{-2}$) & 0.99 & 2.61& 1.29\\
Pauli-$X$ error, $\sqrt{X}$ error  ($10^{-4}$) & 1.89 & 2.1 & 1.74\\
ECR error   ($10^{-3}$) & 0$\_$1& 1$\_$2 &   \\
                            & 9.13 & 6.11&  \\
\end{tabular}
\end{table}

To calculate the error bars presented in Fig. \ref{fig11}, we take into account gate error, readout error, and standard error. The gate error is the sum of the errors of the gates (see Table \ref{trr1}) that define the quantum protocol in Fig. \ref{fig11}, and it is $0.019$.
Note that the estimated value $|\langle U \rangle|^2$ corresponds to the probability of reducing the state $\ket{\tilde{\psi}}$ to the $\ket{000}$ state after measuring three qubits in the standard basis, $|\langle U \rangle|^2 = |\langle{000}|\tilde{\psi}\rangle|^2$. So, the corresponding readout error is the sum of the readout errors of the three qubits. Based on the data presented in Table \ref{trr1}, it is $0.049$.
The standard error is inversely proportional to the square root of the number of shots, $N_\text{shots}$. We estimate the error as $\sqrt{p(1-p)}/\sqrt{N_\text{shots}}$, where $p$ is set to the maximum likelihood estimate of $|\langle U \rangle|^2$, as was done, for instance, in \cite{Suzuki2020}. Considering that $N_\text{shots} = 1024$ and using the maximum likelihood estimate for the results of quantum measurements for each parameter $\phi$ in Fig. \ref{fig11}, we obtained an average standard error of $0.011$.

We fit the obtained result by  $-a\phi^2+b$ with $a$, $b$ being  constants.
Taking into account relation (\ref{rel}),   we find
$\braket{\Delta H_{ch}^2}=aJ^2=4.08J^2$.

\begin{figure}[h!]
		\centering
\includegraphics[scale=0.9]{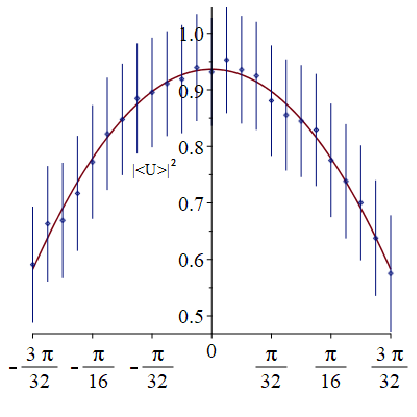}
		\caption{Results of quantum calculations of $|\langle U \rangle|^2$  for spin chain (\ref{hch}) marked by dotes and fitting curve $-4.08\phi^2+0.94$ (line).}.
\end{figure}
On the other hand, using (\ref{hch}) we can also write
\begin{eqnarray}
\braket{\Delta H_{ch}^2}=J^2_{01}+J^2_{12}+2J_{01}J_{12}\braket{\sigma^z_0\sigma^z_2}=
5J^2+4J^2\braket{\sigma^z_0\sigma^z_2}.\label{2h2}
\end{eqnarray}

\begin{figure}[h!]
		\centering
\includegraphics[scale=0.6]{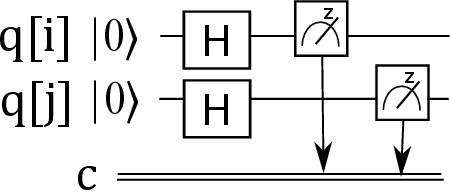}
		\caption{ Quantum protocol for quantifying $\bra{++}\sigma^z_i\sigma^z_j\ket{++}$}
		\label{fig2}
\end{figure}

One can detect the value $\braket{\sigma^z_i\sigma^z_j}=\bra{++}\sigma^z_i\sigma^z_j\ket{++}$ with quantum programming with simple protocol. It is presented in Fig. \ref{fig2}.
 Since the mean value of  $\sigma^z_i\sigma^z_j$ in quantum state $\ket{\psi}$ can be represented as $\bra{\psi}\sigma^z_i\sigma^z_j\ket{\psi}=|\langle{00}|\psi\rangle|^2-|\langle{01}|\psi\rangle|^2-|\langle{10}|\psi\rangle|^2+|\langle{11}|\psi\rangle|^2$
  on the basis of the results of measurements in the standard basis we obtain $\bra{++}\sigma^z_i\sigma^z_j\ket{++}=p_{00}-p_{01}-p_{10}+p_{11}$, where  $p_{00}$,  $p_{01}$,  $p_{10}$,  $p_{11}$ frequencies of quantum states $\ket{00}$, $\ket{01}$, $\ket{10}$, $\ket{11}$ obtained after measurement in the standard basis of two qubits in quantum protocol Fig. \ref{fig2}.

 To calculate $\braket{\sigma^z_0\sigma^z_2}$  the protocol Fig. \ref{fig2} was implemented on qubits $Q_1$ and  $Q_3$ of $\textrm{ibm\_sherbrooke}$  with the number of shots $1024$. On the basis of the results of measurements, we obtain $\braket{\sigma^z_0\sigma^z_2}=0.045$. With regard to the calibration parameters presented in Table \ref{trr1}, the total error of such calculations, which includes the gate error, readout error, and statistical error, is $0.057$.   So, using (\ref{2h2}) for
$\braket{\Delta H_{ch}^2}$ we find $\braket{\Delta H_{ch}^2}=5.18J^2$. Taking into account obtained relation (\ref{h21}) one can also find the sum of the  weighted degrees of nodes in graph $G^{(2)}(V,E)$, we have
\begin{eqnarray}
\sum_in_i^{(2)}=10.36J^2.\label{result_n}
\end{eqnarray}
The theoretical result for this value reads $\sum_in_i^{(2)}=10J^2$.

Similarly for $\braket{\Delta H_{ch}^3}$ and $\braket{\Delta H_{ch}^4}$ we can write
\begin{eqnarray}
\braket{\Delta H_{ch}^3}=(J^3_{01}+3J_{01}J_{12}^2)\braket{\sigma^z_0\sigma^z_1}+(J^3_{01}+3J^2_{01}J_{12})\braket{\sigma^z_1\sigma^z_2}=0.9J,\label{hhhh3}\\
\braket{\Delta H_{ch}^4}=J_{01}^4+J_{12}^4+6J_{01}^2J^2_{12}+4(J^3_{01}J_{12}+J_{01}J_{12}^3)\braket{\sigma^z_0\sigma^z_2}=44.24J^4.\label{hhhh4}
\end{eqnarray}
Now, taking into account the expressions for the curvature and torsion one obtains
\begin{eqnarray}
{\gamma^2\over R^2}=0.649,\ \
{\bar\tau}=0.619\label{result_ct}
\end{eqnarray}
The  results are in agreement with the theoretical ones.  On the basis of analytical calculations for curvature and torsion in this case we have ${\gamma^2/ R^2}={\bar\tau}=0.64$.

\section{Conclusions}

Weighted graph states of the spin system described by Ising model with anisotropic interaction have been studied. The velocity, curvature, and torsion of the states have been calculated with the usage of their relation with the fluctuations of energy. We have found that the velocity of quantum evolution is determined by the sum of the weighted degrees of nodes in the graph $G^{(2)}(V,E)$.  The graph $G^{(2)}(V,E)$ is constructed by squaring of the wights of graph $G(V,E)$ that represents the weighted graph state (\ref{velocity1}). The
 curvature (\ref{res2}) depends on the sum of products of weights of edges creating squares in graph $G(V,E)$  and also on the sum of the weighted degrees of nodes in the graphs $G^{(2)}(V,E)$, $G^{(4)}(V,E)$. The graph $G^{(4)}(V,E)$ can be obtained by  rising to the fourth the wights of graph $G(V,E)$.  The torsion in addition depends on the sum of the weighted degrees of nodes in the graph $G^{(3)}(V,E)$ constructed by cubing of the wights of graph $G(V,E)$. The torsion is also determined by the sum of products of weights of edges in graph  $G(V,E)$ creating triangles $S_3$ (\ref{res3}).

So,  we have established a relationship between the geometrical properties of evolutionary quantum graph states and the properties of the corresponding graphs.  It is not a simple task to compute the sum of the products of the weights of edges  that form triangles $S_3$ or squares $S_4$ in the case of large graph (in the case of unweighted graph, these values are related to the number of triangles and the number of squares). The obtained results for
$\langle(\Delta H_I)^3\rangle$ (\ref{h32}), $\langle(\Delta H_I)^4\rangle$ (\ref{dh41}) open the possibility to detect these properties of a graph on a quantum device. It is worth  mentioning that $\langle(\Delta H_I)^3\rangle$, $\langle(\Delta H_I)^4\rangle$
can be detected through quantum programming based on studies of the mean value of the evolution operator $|\langle U \rangle|^2$ at small times, considering terms in the expansion (\ref{rel}) up to $t^6$. For such studies, the development of quantum processors with reduced quantum errors is essential.
In this paper, as an example, and to minimize the errors of quantum calculations, we studied a simple three-spin system (a spin chain). For this case, we studied $|\langle U \rangle|^2$ on a quantum device and detected $\langle(\Delta H_I)^2\rangle$. For such a simple example, the mean values $\langle(\Delta H_I)^2\rangle$, $\langle(\Delta H_I)^3\rangle$, $\langle(\Delta H_I)^4\rangle$,  can be easily calculated analytically or quantified based on studies of $\braket{\sigma^z_0\sigma^z_2}$
on a quantum device (see equations (\ref{2h2}), (\ref{hhhh3}), and (\ref{hhhh4})), as was also demonstrated in the paper.

  Geometric properties of weighted graph states have been studied  with quantum programming on the basis of their relation with the fluctuations of energy. In the particular case of a spin chain (\ref{hch}) we have calculated the curvature and torsion and also the sum of the weighted degrees of nodes in the graph $G^{(2)}$  on the basis of results of  programming on IBM's quantum computer $\textrm{ibm\_sherbrooke}$. The results obtained on the basis of quantum calculations (\ref{result_n}), (\ref{result_ct}) are in agreement with the theoretical ones.

 We hope that with the development of precise multi-qubit quantum processors, the obtained results will enable the quantum computation of the properties of large graphs and open the way for achieving quantum supremacy in such studies.

\section*{Acknowledgment}

This work was supported by the Virtual Ukrainian Institute of Advanced Studies. The author thanks Prof. Tkachuk V. M. for many useful comments and support during the research studies

\section*{Research Data Policy and Data Availability Statements}
The datasets generated during and/or analyzed during the current study are available from the corresponding author
on reasonable request.



\end{document}